# Exploring the transfer of plasticity across Laves phase interfaces in a dual phase magnesium alloy

Julien Guénolé[a,b,c], Muhammad Zubair[a,d], Swagata Roy[a], Zhuocheng Xie[a], Marta Lipińska-Chwałek[e,f], Stefanie Sandlöbes-Haut[a], Sandra Korte-Kerzel[a]

[a] Institute of Physical Metallurgy and Materials Physics, RWTH Aachen University, 52056 Aachen, Germany.
[b] Université de Lorraine, CNRS, Arts et Métiers ParisTech, LEM3, 57070 Metz, France
[c] Labex Damas, Université de Lorraine, 57070 Metz, France.
[d] Department of Metallurgical and Materials Engineering, G.T Road, UET Lahore, Pakistan.
[e] Central Facility for Electron Microscopy, RWTH Aachen University, 52074 Aachen, Germany.
[f] Ernst Ruska-Centre for Microscopy and Spectroscopy with Electrons, Forschungszentrum Jülich GmbH, 52425 Jülich, Germany.

**Abstract**

The mechanical behaviour of Mg-Al alloys can be largely improved by the formation of an intermetallic Laves phase skeleton, in particular the creep strength. Recent nanomechanical studies revealed plasticity by dislocation glide in the $(Mg,Al)_2Ca$ Laves phase, even at room temperature. As strengthening skeleton, this phase remains, however, brittle at low temperature. In this work, we present experimental evidence of slip transfer from the Mg matrix to the $(Mg,Al)_2Ca$ skeleton at room temperature and explore associated mechanisms by means of atomistic simulations. We identify two possible mechanisms for transferring Mg basal slip into Laves phases depending on the crystallographic orientation: a direct and an indirect slip transfer triggered by full and partial dislocations, respectively. Our experimental and numerical observations also highlight the importance of interfacial sliding that can prevent the transfer of the plasticity from one phase to the other.

# 1. Introduction

The mechanical response of composite materials is dependent on the properties of their different phases individually, but also largely on the properties of their interfaces. Lightweight Mg-based composites with a controlled proportion of Al and Ca can be strengthened by the precipitation of a Laves phases skeleton [1]. The reinforcement potential of Laves phases in these alloys is primarily due to their high strength compared to the matrix phase [2]. Laves phases are hard intermetallic phases with a topologically close packed structure arranged in a cubic (C15) or hexagonal (C14 and C36) unit cell [3,4]. The mechanical response of such a dual phase microstructure typically shows plasticity in the Mg matrix and fracture in the hard-intermetallic phase [1,5]. While effectively strengthening the composite, the brittleness of the Laves phase limits the maximum strength and the formability of the composite.

Plasticity in hexagonal Mg-based Laves phases has been observed experimentally at room temperature by means of nano-mechanical testing [2]. In particular, $Mg_2Ca$ (C14) micro-pillar compression revealed the activation of basal, prismatic and pyramidal slip systems [2]. Additionally, atomistic simulations with the same phase confirmed that synchroshear dislocations are activated for basal slip [6]. To our best knowledge, one study investigated the properties of interfaces in the similar Mg-$Zn_2$Mg composites [7], and none investigated the transfer of plasticity from the matrix to the Laves phase.

Different Laves phases exhibit different dislocation pinning characteristics as observed for Laves phase reinforced Ti alloys [8]. In case of Mg alloys, Laves phases and their interfaces are strong obstacles for dislocation motion, making Mg alloys containing Laves phases well suited for high temperature applications in which creep resistance is required [9]. However, a lack of slip transfer from α-Mg to Laves phases also results in limited plasticity as displayed by the low ductility of these alloys at room and elevated temperatures [1]. Easier slip transfer is usually associated with more homogenous plastic flow [10] and in case of Mg-Al-Ca alloys, this would presumably result in better bulk plasticity at the expense of creep resistance. The knowledge of slip transfer mechanisms may thus prove essential in designing alloys with a tailored balance between ductility, strength and creep resistance.

Here, we explore the co-deformation of Laves phases with an α-Mg matrix by the combination of experimental and numerical approaches. The surface morphology of a deformed composite is analysed by scanning electron microscopy. Cracks and slip traces are characterized in both, the matrix and the Laves phase skeleton. In parallel, atomistic simulations are performed for a similar composite and several possible orientation relationships are investigated. Numerical nano-mechanical tests and controlled dislocation-interface interactions at the atomic scale enlighten the experimental results. Finally, we show mechanisms of transfer of plasticity from the Mg matrix to the Laves phase skeleton across the interface.

## 2. Experiments

A cast Mg-5.21Al-3.18Ca (wt.%) alloy was prepared from raw materials consisting of pure Mg, Ca and Al. The raw materials were molten in an induction furnace and solidified under 15 bar Ar pressure in a steel crucible. The chemical composition was determined through wet chemical analysis.

The samples for microstructural examination were first ground using 4000 SiC emery paper, followed by mechanical polishing using 3 and 1 µm diamond suspension. The mechanically polished samples were then subjected to electrolytic polishing using the AC-II (Struers) electrolyte. Electro-polishing was carried out at ≤ -20 ºC, 15V and for 60 s. Different electrochemical behaviour of the metallic and intermetallic phases resulted in the appearance of a waviness on the sample surface which was then removed by fine polishing with OPU (≈ 40 nm $SiO_2$ colloidal suspension). After fine polishing, the samples were cleaned in an ultrasonic bath followed by light pressing on gently rotating cleaning cloth containing ethanol as a cleaning agent/lubricant. The composition of the α-Mg matrix and intermetallic phase was determined with the aid of energy dispersive X-ray spectroscopy (EDS) performed at an accelerating voltage of 10kV in SEM. This relatively low voltage was selected to limit the interaction volume.

Standard dog-bone shaped specimens with a gauge length of 10 mm were deformed in an electromechanical testing machine (ETM) at 170 °C in tensile mode at a constant strain rate of $5 \times 10^{-4}$ $s^{-1}$. The specimens were later characterized in a scanning electron microscope (SEM; Zeiss LEO1530) using secondary electron (SE) imaging and electron backscatter diffraction (EBSD) after pre-determined deformation steps of 3 and 5%. An acceleration voltage of 20 kV was used for EBSD and 10-20 kV for SE imaging. The EBSD data were analysed using Channel 5 and Matlab 2018b software. A step size in between 0.25 and 0.7 µm was used for EBSD depending on the scanned area. Focussed Ion beam (FIB) milling (in FEI Helios Nanolab 600i) was used to prepare the site-specific electron transparent samples (FIB lamella) for investigations in transmission electron microscope (TEM). The FIB lamella cut plane was selected to be perpendicular to the polished surface of the bulk sample and to the tensile direction, as shown in Figure 2. The orientation relationship between matrix and intermetallic phase and EDS analysis of their chemical composition were performed by TEM (FEI Tecnai G2 F20 [11] ) at an accelerating voltage of 200 kV.

### 2.1. Experimental results

Figure 1 represents an SE image taken from a sample deformed to 3% global strain. It can be seen in the figure that the microstructure consists of two phases, i.e. α-matrix reinforced with an intermetallic Laves phase skeleton. Note that grains are much larger than the cell size or, in other words, phase boundaries are not necessarily grain boundaries (See supplementary material, Figure SM1). As a consequence, to a scale far beyond that studied and presented in Figure 1, the sample is a single crystal (See supplementary material, Figure SM2). A comparative observation before and after deformation at 170°C reveals no modification in the morphology of the Laves phase network induced by, for example, the temperature (See supplementary material, Figure SM3).

The composition of the intermetallic phase as determined from EDS point analysis performed on the intermetallic phase (at 8 points in SEM) is presented in the . The EDS point analysis confirms the presence of all three elements in significant quantities in the Laves phase and is considered to be C36 ((Mg,Al)$_2$Ca) phase. Similar compositions for C36 Laves phase via EDS have also been reported earlier [12,13]. The α-Mg matrix composition was measured at three (3) different points [22]. Further, EDS was also performed in TEM (to measure the Laves phase separately from the matrix) and the results for both phases present in TEM lamella (shown in Figure 2) are given in Table 1.

*Table 1: Composition of intermetallic and matrix phases as determined from EDS in SEM and TEM.*

| Microscope | | Mg (at. %) | Al (at. %) | Ca (at. %) | Remarks |
| --- | --- | --- | --- | --- | --- |
| SEM | Intermetallic phase | 49.7 ± 11.3 | 31 ± 6.4 | 19.3 ± 4.9 | C36 phase |
| SEM | α-Mg matrix | 98.3 | 1.7 | - | Solid solution |
| TEM | Intermetallic phase | 25.5 ± 3.8 | 45.5 ± 4.1 | 29.0 ± 3.2 | C36 phase |
| TEM | α-Mg matrix | 98.2 ± 0.5 | 1.7 ± 0.4 | 0.14 ± 0.07 | Solid solution |

Moreover, an EDS area scan was performed in the SEM on the same alloy and is presented in the supplementary data (Figure SM4). From this figure, it is clear that the alloying elements are mainly concentrated in the intermetallic phase. There is negligible amount of Ca in the α-Mg matrix (Table 1) because of limited solubility of Ca in α-Mg [14].

As can be seen from the comparison of the compositions determined from EDS in SEM and TEM (see ), the influence of the matrix is diminished in case of EDS in TEM. However, there still is a considerable amount of Mg present in the intermetallic phase showing that it is indeed C36 Laves phase. This argument is further validated from the respective SAED pattern shown in Figure 2(e).

The same intermetallic phase was also reported by several other researchers in similar alloys, i.e. Suzuki et al. [15] in the Mg-5Al-3Ca alloy and by Luo et al. [16] in AC53 (Mg-4.5Al-3Ca-0.27Mn) alloy. They used TEM to confirm their results. In our results, slip lines in the matrix phase are clearly visible and highlighted by blue arrows. Using EBSD-assisted slip line analysis, these slip lines were identified as basal slip traces in the α-Mg matrix. At points where the slip lines interact with the (Mg,Al)$_2$Ca (C36) Laves phase, cracks in the C36 phase can be observed. This is in agreement with our earlier work [1].

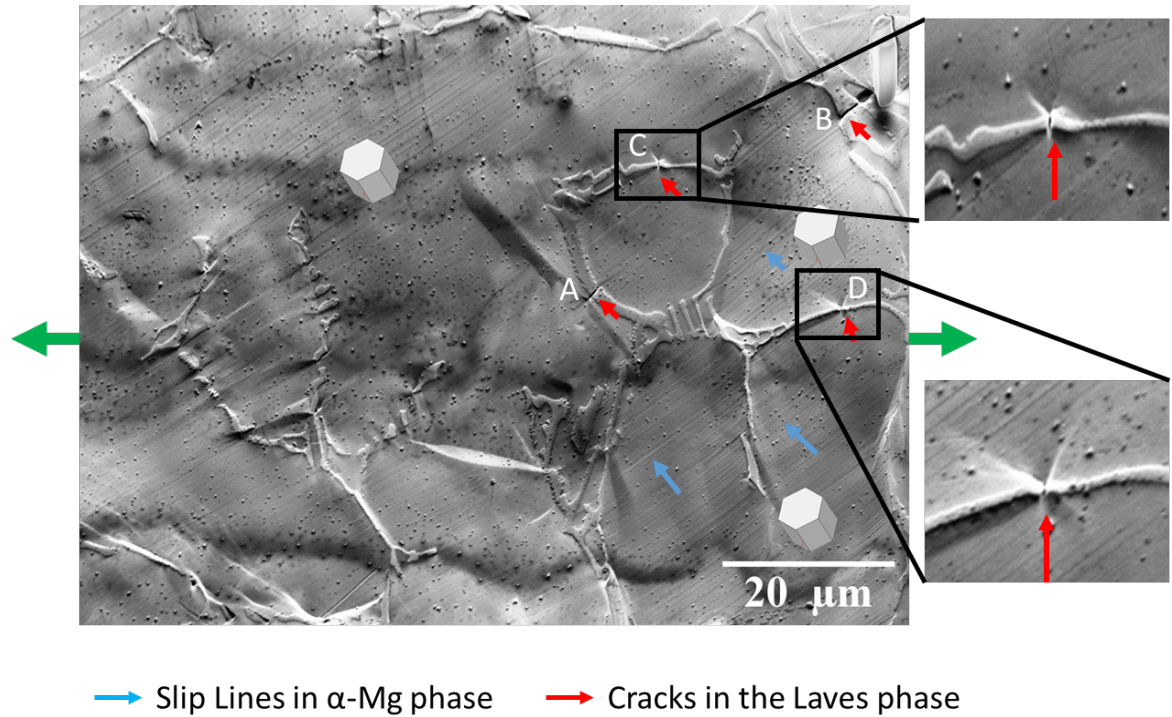

→ Slip Lines in α-Mg phase   → Cracks in the Laves phase

*Figure 1: SE image (with the stage tilted 70 ° towards the electron beam, image tilt-corrected) of a sample deformed to 3% at 170 °C. Green arrows represent the direction of the applied force, blue arrows show basal slip lines in the α-Mg phase and red arrows indicate cracks in the Laves phase. The insets show the magnified images of the cracks in the C36 phase which are perpendicular to the loading direction at points C and D. Unit cells depict the orientation of α-Mg matrix.*

Figure 2 reveal co-deformation of the α-Mg matrix and the intermetallic C36 Laves phase in the sample deformed to 5% global tensile strain at ≈170 °C. The slip lines in the α-Mg and C36 phases are highlighted with blue and orange arrows, respectively. Unfortunately, it was very difficult to analyse the orientation of the C36 Laves phase from EBSD data. This is primarily because of the small size of the intermetallic phase in comparison to the α-Mg matrix and also their structural similarity with the hexagonal Mg matrix. Therefore, a detailed TEM analysis was performed on the FIB lamella lifted out of the material volume indicated in Figure 2 (b) with white dotted rectangle. TEM investigations yielded information about the orientation relationship between the Mg matrix and the C36 Laves phase and confirmed the presence of a basal slip deformation mode in the Laves phase. Figure 2 (c-g) show microstructure details of the deformed material, as observed with the aid of TEM. Planar defects corresponding to the basal slip events were observed in α-Mg matrix, as well as in the intermetallic phase. In particular, bright field (BF) TEM images in Figure 2 (d) and (f) show basal slip planar defects in either phase observed in edge-on orientation (i.e. slip plane is parallel to the viewing direction). Accordingly, planar defects have appearance of thin lines passing diagonally the BF images (indicated with blue arrows in α-Mg matrix (d) and with orange arrows in the Laves phase (f)). In either case, the observation direction is parallel to the respective [2 $\bar{1}$ $\bar{1}$ 0] crystallographic direction. Corresponding SAED patterns are given in Figure 2 (e) and (g), respectively. It is worth to note, however, that different sample orientations were required for [2 $\bar{1}$ $\bar{1}$ 0] zone orientations of each phase. A detailed analysis of the crystallographic orientation relationship between α-Mg matrix and the Laves phase revealed that basal planes of both phases are almost perpendicular to each other (∠[0 0 0 1]$_{Mg}$, [0 0 0 1]$_{C36}$ = 79° and ∠[2 $\bar{1}$ $\bar{1}$ 0]$_{Mg}$, [1 0 $\bar{1}$ 0]$_{C36}$ = 8°). Similar ORs (∠[0 0 0 1]$_{Mg}$, [0 0 0 1]$_{C36}$ = 82° ± 10°) were systematically observed also for different FIB lamellas cut out of other sample grains (not described here), indicating the tendency of the composite alloy to adopt a preferential crystallographic orientation relationship, with basal planes of constituent hexagonal phases being almost perpendicular to each other.

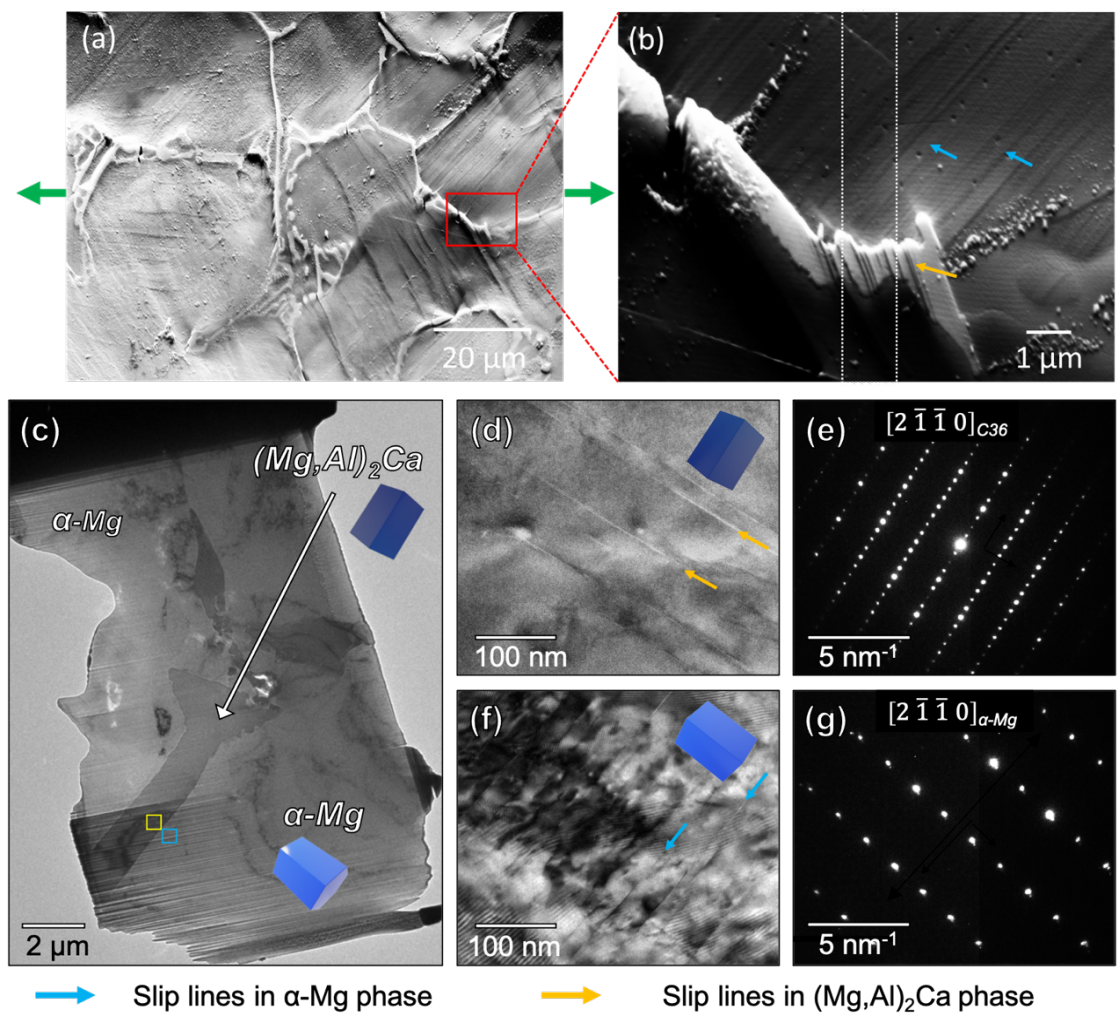

*Figure 2: Microstructure of the sample deformed to 5% at 170°C: (a) SE SEM overview image (stage tilted 70° to the electron beam, image tilt-corrected) with green arrows indicating the direction of the tensile force applied in deformation test. (b) Magnified image (SE SEM) of the microstructure marked in (a) with red rectangle. The white dotted rectangle in (b) indicates, in top view, the position of the FIB cut. (c) Overview (BF TEM) image of the FIB lamella in the initial orientation (no sample tilt). Yellow and blue rectangles indicate sample areas shown in detail in (d) and (f), respectively. (d-g) BF images of the basal-slip structures observed in (Mg,Al)$_2$Ca (d) and α-Mg matrix (f), with*

*corresponding [2 $\bar{1}$ $\bar{1}$ 0] SAED patterns in (e) and (g). Unit cell projections in (c), (d) and (f) schematically indicate orientation of the hexagonal structures in the respective images (bright blue for α-Mg and dark blue for (Mg,Al)$_2$Ca).*

## 2.2 Discussion on the experiments

It can be observed in Figure 1 that the cracks in the C36 Laves phase occur at points where slip lines in the α-Mg matrix interact with the Laves phase. Further, it can be seen in Figure 1 that the cracks at points A and B are oriented around 45-50 ° to the globally applied tensile stress direction. Moreover, these cracks in the C36 phase are almost parallel to the basal slip lines in the α-Mg matrix. This indicates that the cracks are developed due to the shearing of the C36 Laves phase by basal slip in the α-Mg matrix. On the other hand, the cracks at points C and D (as can be seen enlarged in the insets of Figure 1) are almost perpendicular to the applied stress direction and correspond probably to brittle fracture of the C36 phase. Figure 2, on the other hand, shows an event where plastic flow is actually transmitted from the α-Mg matrix into the Laves phase. Basal defect structures are indeed observable in the BF TEM micrographs of both phases, represented in Figure 2 (d and f). This suggests that in addition to α-Mg matrix, the intermetallic Laves phase (C36) is indeed plastically deformable at length scales present in the as-cast Mg-5.21Al-3.18Ca alloy. The α-Mg matrix offers significantly lower resistance to dislocations movement as compared to the Laves phases. Specifically, the critical resolved shear stress (CRSS) values for basal, prismatic and pyramidal slip in Mg amounts to ≈ 0.52 MPa [17], ≈ 39 MPa [18] and ≈44 MPa [19], respectively (as determined by macroscopic testing of pure Mg, except in case of [19] where the CRSS for pyramidal slip was determined through micropillar compression). The CRSS values for the Mg phase are much lower than the CRSS values reported for the same slip systems in the C14 Mg$_2$Ca Laves phase, which were determined as ≈520 MPa (basal slip), ≈440 MPa (prismatic slip) and ≈530 MPa (pyramidal slip) by micropillar compression [2] (NB: a direct comparison is impeded by the brittleness of the Laves phase and the size effect encountered in microcompression testing of Mg, resulting in a CRSS for basal slip of the order of ≈ 7 MPa as opposed to 0.52 MPa [17] measured macroscopically [20]). The other Laves phase (C15 Al$_2$Ca phase) in the Mg-Al-Ca system was reported to be even harder than the C14 Mg$_2$Ca Laves phase [21]. As the hexagonal C36 Laves phase is very closely related to the hexagonal C14 Laves phase, we assume here that plastic deformation follows the same mechanisms. Owing to this difference in CRSS; it appears likely that under the application of a tensile stress, slip in the α-Mg matrix is activated first and then at points of severe stress concentrations, for example where slip lines in the Mg phase interact with the Laves phase, plastic flow can be transmitted into the Laves phase. Further, it is possible that at those intersection points of slip in the Mg matrix with the Laves phase, where cracks were observed, some plastic flow has occurred prior to cracking. However, it is not possible to distinguish cracking with and without prior slip in the Laves phase by considering only post-mortem results.

As reported elsewhere, it is worth mentioning that considerable strengthening is induced by the Laves phases, i.e. the yield and tensile strength of this alloy in the as-cast state is 92 ± 2MPa and 123 ± 11MPa, respectively [22] at room temperature, whereas the same values for pure Mg in the as-cast state are 26.6 and 69.2 MPa, respectively [23].

## 3. Simulations

To explore the atomic-scale details of the slip transfer observed experimentally and to understand the underlying mechanisms, we performed atomistic simulations with semi-empirical potentials. As no reliable potential is available to date that correctly models the plasticity in the ternary Mg-Al-Ca intermetallic phase, we consider here the C14 $Mg_2Ca$ Laves phase as surrogate for the C36 $(Mg,Al)_2Ca$ Laves phase. Similar to previous experimental [16] and numerical work [7], two different orientation relationships (OR) have been considered in our numerical approach. One with both phases' basal planes parallel, $(0001)_\alpha \parallel (0001)_{Mg_2Ca}$, the other with the basal plane of the Mg phase aligned with the prismatic plane of the $Mg_2Ca$, $(0001)_\alpha \parallel (11\bar{2}0)_{Mg_2Ca}$, close to the OR we observed by TEM.

### 3.1. Numerical methods

Atomistic simulations have been carried out with the classical molecular dynamics code LAMMPS [24]. Interatomic interactions were modelled by the recently developed potential of Kim et al. [25] based on the modified embedded atom method (MEAM). This potential has been specifically optimized to represent both, hexagonal Mg phase and the C14 $Mg_2Ca$ Laves phase, and it has been used successfully to model the complex dislocation motion in the basal plane of the C14 Laves phase [6]. This work involves molecular static (MS) and molecular dynamic (MD) simulations. MS simulations were performed by using both conjugate gradient [26] and FIRE [27,28] algorithms, and the configurations were considered optimized for a norm of the global force vector below $10^{-8}$ eV/Å. MD simulations were performed within either the microcanonical (NVE, $T_0$ = 0K) or the canonical (NVT, $T_0$ > 0K) thermodynamic ensemble with a timestep $\Delta t$ = 1.0 fs. In the NVT ensemble, the temperature was controlled by a Nosé-Hoover thermostat [29] with a damping parameter of 0.1 ps. The atomistic configurations were constructed using Atomsk [30], and visualised and analysed using OVITO [31]. The defects in both, Mg and $Mg_2Ca$ phases were evidenced by calculating atomic-level strain tensors at each particle. More specifically, the von Mises local shear invariant, $\gamma_{vM}$, was estimated from the atomic deformation gradient tensor calculated based on atom displacements [32]. Note that the commonly used structural identification methods, like the common neighbours analysis (CNA) [33], are not adapted to the C14 structure and defects at complex interfaces have been already characterized by such shear invariant based method [34].

### 3.2. Numerical results

#### 3.2.1. Parallel basal planes: $(0001)_\alpha \parallel (0001)_{Mg_2Ca}$

We first focus on the OR $(0001)_\alpha \parallel (0001)_{Mg_2Ca}$, which forms a composite that exhibits coplanar basal planes. As the objective of the simulation is to study the transfer of slip at the interface from the Mg matrix, the onset of plasticity in the Mg phase is favoured by choosing an angle of 45° with the deformation axis. This maximises the resolved shear stress on the primary (basal) slip planes. The corresponding atomistic sample shown in Figure 3 is made of Mg and Mg₂Ca phases separated by an interface perpendicular to the Z-axis, and contains 996,941 atoms. The phases are oriented as follows: X'-$[\bar{1}100]$, Y'-$[11\bar{2}0]$, Z'-$[0001]$ for Mg, and X''-$[2\bar{1}\bar{1}0]$, Y''-$[\bar{1}2\bar{1}0]$, Z''-$[0001]$ for Mg₂Ca. The crystallographic directions X' and X'' are aligned with the box axis X. The crystallographic directions Z' and Z'' are parallel and have an angle of 45° with the Y axis. No periodic boundary conditions are used. Uniaxial deformation is dynamically applied along the Y-axis: the sample is homogeneously scaled at a strain rate of $10^8 s^{-1}$ and the deformation is maintained by means of 2D boundary conditions. The deformation was performed at T≈50K within the microcanonical ensemble ($T_0$=0K). In this case, the temperature T corresponds to the average temperature during the simulated deformation. Tests at 500K and 700K have been also performed within the canonical ensemble.

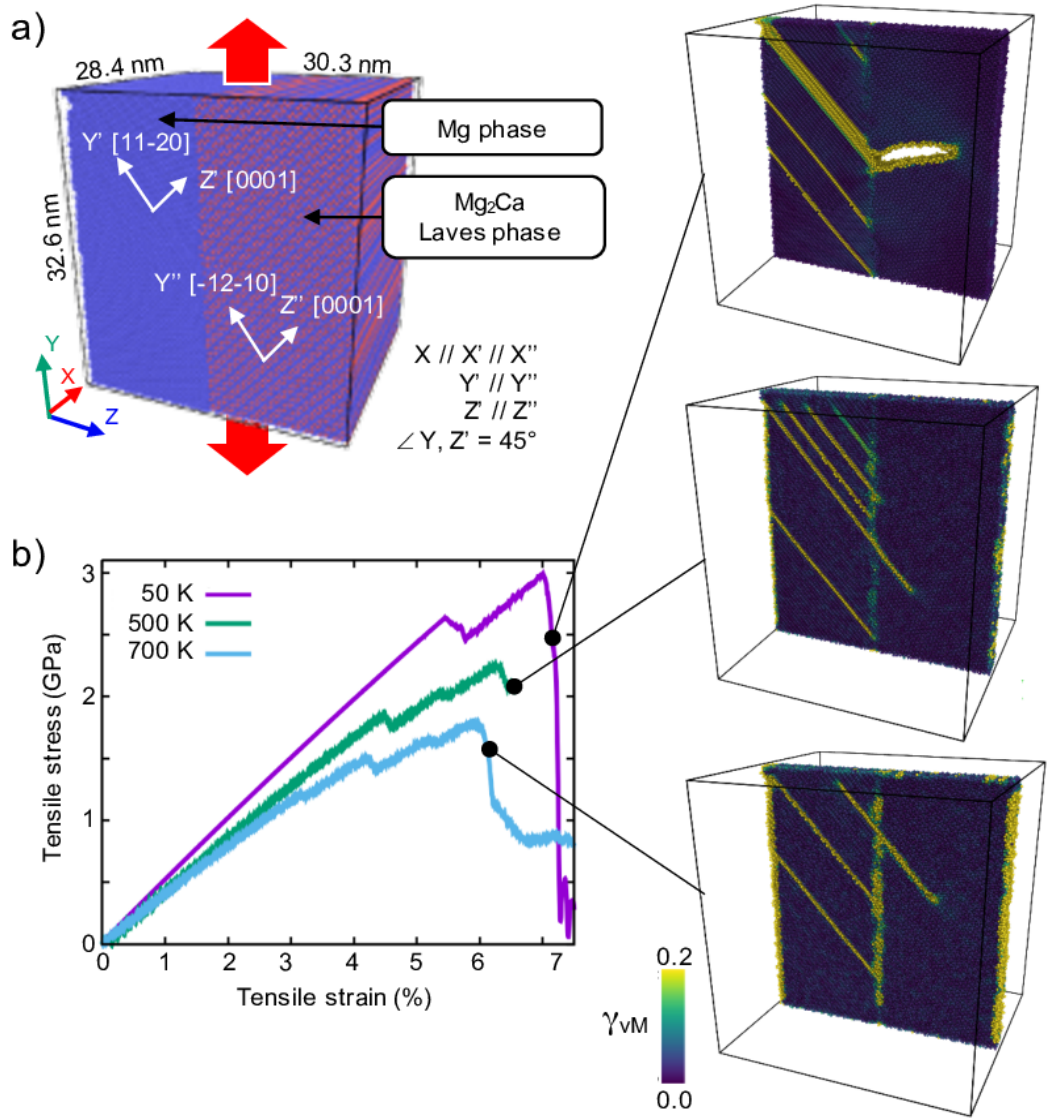

*Figure 3: Direct slip transfer at an Mg-Mg$_2$Ca interface during nanomechanical tensile tests. **a)** Simulation setup with the orientation relationship OR $(0001)_\alpha \parallel (0001)_{Mg_2Ca}$. Red arrows indicate the pulling direction. Atoms in blue (red) are Mg (Ca). **b)** Stress-strain curves of the tensile tests at 50K (purple), 500K (green) and 700K (blue). Insets show sliced snapshots of the first plastic event in the Mg$_2$Ca phase. Atoms are coloured according to their averaged local shear strain $\gamma_{vM}$.*

*Figure 3*b shows the stress-strain curves of our numerical strain-controlled nanomechanical tensile tests at 50K, 500K and 700K. Note, that we also performed compression tests on the same sample, and they produced comparable results. All three curves exhibit a similar profile. First, there is an initially purely elastic co-deformation regime: both phases deform with negligible plasticity limited to surfaces. Then small stress drops indicate the activation of plasticity in the Mg phase by means of dislocation nucleation and glide from the surface or from the interface. Finally, a large stress drop corresponds to the initiation of either fracture at low temperature or slip at high temperature in the Mg$_2$Ca phase, as illustrated by the simulation snapshots in *Figure 3*b. In all cases, fracture or plasticity in the Mg$_2$Ca phase are initiated from locations where Mg basal slip lines interact with the interface. A closer look at the mechanism is provided in *Figure 4* (See also Movie S1), which shows a sequence of snapshots of the sample deformed at 500K. The slip mechanisms are highlighted in both phases by the calculation of the local shear strain, or von Mises local shear invariant, as described above. At ε = 3.08%, two slip traces are visible. The stronger one indicates that more than one dislocation has glided within the same glide plane. At ε=5.8%, 3 more slip traces are visible with a local shear strain concentration at the location of slip transfer, as highlighted in the inset. At ε=5.9%, no additional slip traces appear in the Mg phase, but the slip trace in the Mg phase highlighted in red shows a higher shear strain level indicating additional dislocation glide. As a consequence, a slip event in the $Mg_2Ca$ phase is initiated as highlighted in the magnified views (*Figure 4*). At ε=6.0%, it propagates further within the basal plane of the $Mg_2Ca$ phase by showing similarities with the synchroshear mechanism, typical for basal slip in Laves phases [6,35].

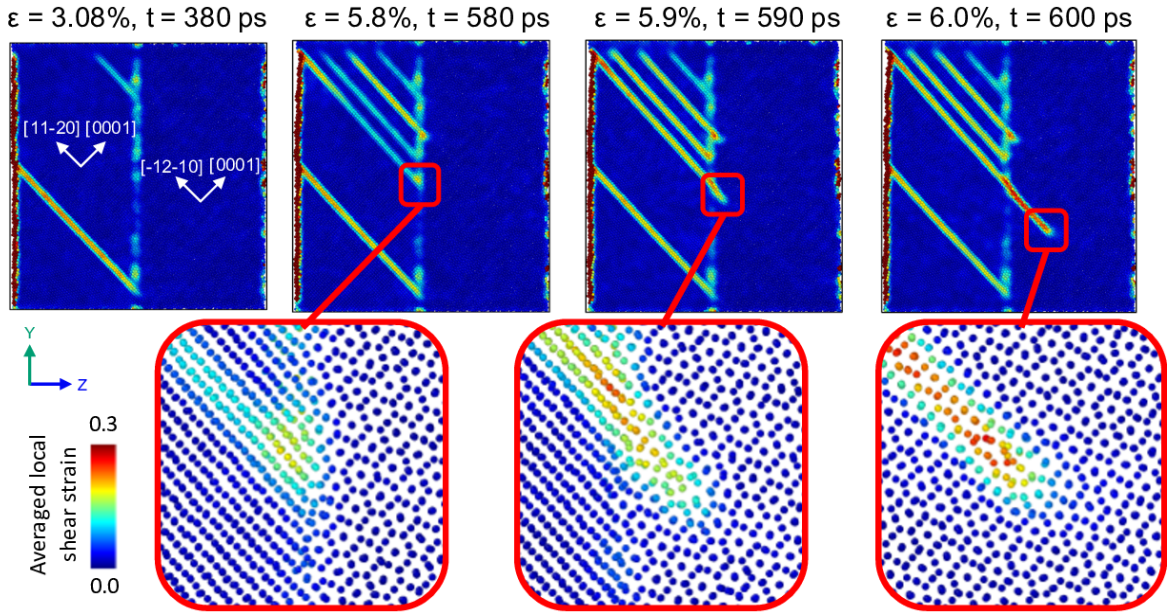

*Figure 4: Details of direct slip transfer at an Mg-$Mg_2Ca$ interface $(0001)_\alpha \parallel (0001)_{Mg_2Ca}$ during nanomechanical tensile tests at 500K. Atoms are coloured according to their averaged local shear strain $\gamma_{vM}$. Insets show the details of the slip transfer at the interface.*

Our simulation at low temperature (50K) does not show any slip within the $Mg_2Ca$ phase. We reduced the deformation rate up to 2 orders of magnitude without preventing the formation of a crack from the interface and through the $Mg_2Ca$ phase.

### 3.2.2. Non-parallel basal planes: $(0001)_\alpha \parallel (11\bar{2}0)_{Mg_2Ca}$

The particular orientation relationship investigated in the previous section with the basal planes aligned parallel in both phases obviously favoured basal-to-basal slip transfer at the interface. To obtain some insights into the possibility of having such a slip transfer for other orientation relationships, we considered the OR $(0001)_\alpha \parallel (11\bar{2}0)_{Mg_2Ca}$, which is similar to the one observed in our experiments and in the literature [7]. In this configuration, the basal plane of the Mg phase is aligned parallel with a prismatic plane of the Mg2Ca phase. The corresponding atomistic sample shown in Figure 5a is composed of the Mg and Mg$_2$Ca phases separated by an interface normal to the Z-axis, and contains a total of 147,060 atoms. The phases are oriented as follows: X'-[1$\bar{1}$00], Y'-[0001], Z'-[11$\bar{2}$0] for Mg, and X''-[1$\bar{1}$00], Y''-[11$\bar{2}$0], Z''-[0001] for Mg$_2$Ca. The crystallographic directions Z' and Z'' are aligned with the box axis Z. The crystallographic directions Y' and Y'' are parallel and have an angle of 45° with the Y axis. No periodic boundary conditions are used. Uniaxial compression is dynamically applied along the Y-axis: the sample is homogeneously scaled, and the compression is maintained by means of a force wall defined by a harmonic potential. The quasi-static deformation test was performed by successive MS simulations. The deformation at T≈50K was performed within the microcanonical ensemble (T$_0$=0K). A test at 300K has been performed within the canonical ensemble. Both dynamic deformations were carried out at a strain rate of 10$^8$/s.

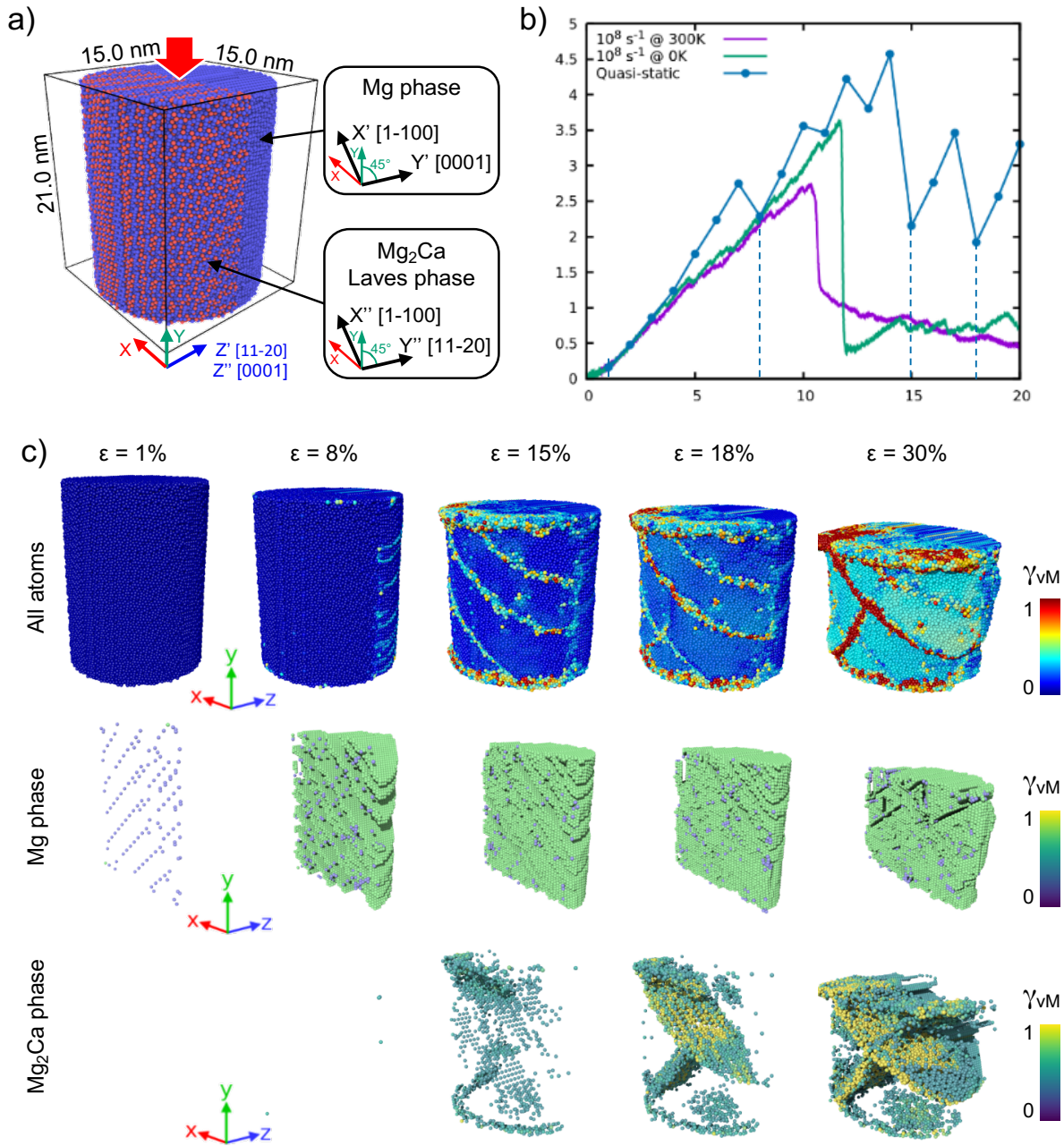

Figure 5: Indirect slip transfer at the Mg-Mg$_2$Ca interface during nanomechanical compression tests. **a)** setup of the compression test simulation, with the orientation relationship $(0001)_\alpha \parallel (11\bar{2}0)_{Mg_2Ca}$. Red arrows indicate the compression direction. Atoms in blue (red) are Mg (Ca). **b)** stress-strain curves of the compression tests. The green and purple solid lines correspond to dynamic deformation at ≈50K and 300K. The dots of the blue line correspond to the relaxed states of the quasi-static deformation. Vertical dashed lines are guide for the eyes at strains 1%, 8%, 15% and 18% that correspond to the snapshots shown in c. **c)** snapshots of the quasi-static deformation test. First row: all atoms are shown. Second (third) row: only atoms of the Mg (Mg$_2$Ca) phase with an atomic shear strain γ$_{vM}$ greater than 0.3 are shown. Atoms are coloured according to the atomic shear strain γ$_{vM}$.

Figure 5b shows the stress-strain curves of the three compression tests. The quasi-static deformation shows a profile that is more staggered than the dynamic deformations, but all tests follow a similar trend. To describe the mechanisms in detail, we will focus on the quasi-static deformation. Figure 5c shows sequences of snapshots of the quasi-static deformation from 1% to 30% compressive strain. The first row shows the entire sample during deformation, evidencing that slip traces form at the surface (See also Movie S2). The second and third rows show only those atoms that have an atomic von Mises shear invariant $\gamma_{vM}$ larger than 0.3. In other words, slip activities are highlighted by hiding atoms that have not undergone significant shear. As a lot of events happened during the deformation, we split the sample in two parts to ease the visualisation: the second and third rows show the Mg and $Mg_2Ca$ phases only, respectively (see also Movies S3 and S4). Up to 8% strain, the deformation is purely elastic and no slip events are detected. At 8% strain, a slight drop of the stress occurred (first dashed vertical line in *Figure 5*b) as the Mg phase shows a strong plastic activity with the nucleation of dislocations and the formation of twins. The $Mg_2Ca$ phase is deformed elastically up to 15% strain as the Mg phase continues to deform plastically. At 15% strain, shear events are triggered in the $Mg_2Ca$ phase, principally on two approximately orthogonal planes. The shear continues to increase while deforming the sample up to 30%. Note that while the Mg phase appears almost entirely twinned, most of the plastic shear in the $Mg_2Ca$ phase is concentrated in the two planes initially activated at 15% strain.

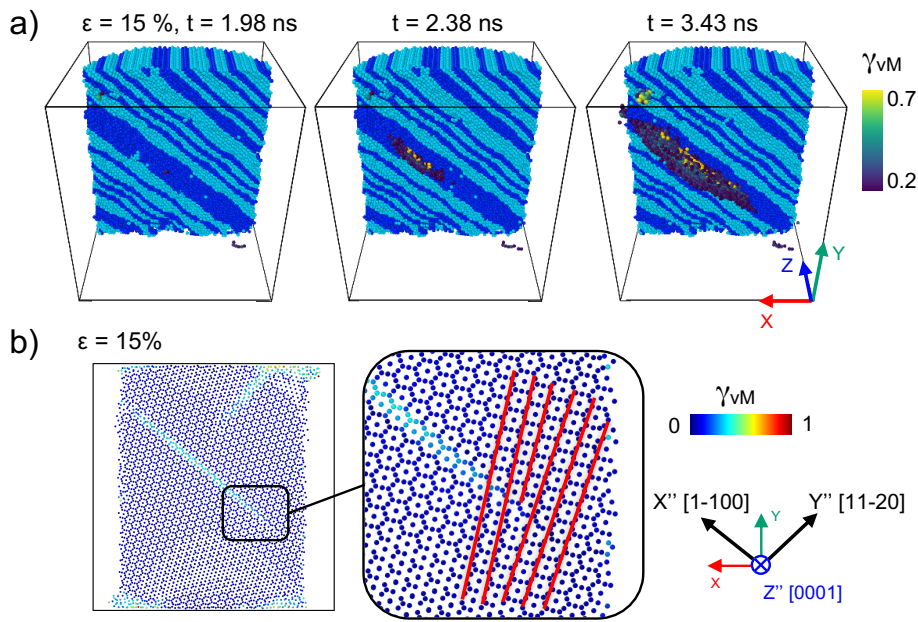

*Figure 6: **a)** snapshots of the dynamic deformation at ≈50K showing the nucleation of a dislocation in the Laves phase from the interface. Mg phase: all atoms visible, coloured according to their crystal structure (HCP, light blue; FCC, dark blue). $Mg_2Ca$ phase: only atoms with an atomic shear strain greater than 0.3 are visible. **b)** sliced view of the $Mg_2Ca$ phase deformed to 15% during a quasi-static simulation. The inset highlights the non-dissociated core of the prismatic dislocation nucleated from the interface in b). Atoms are coloured according to the atomic shear strain.*

A deeper insight in the onset of plasticity in the $Mg_2Ca$ phase is given by the kinetics of the mechanism from the dynamic simulation, as shown in Figure 6. Here, the Mg phase appears already largely twinned at a compressive strain of 15% and a time t = 1.98 ns. While maintaining the strain constant, the dynamics is simulated further (snapshots at t = 2.38 ns and t = 3.43 ns, Figure 6a and Movie S5) and evidences the onset of slip at the Mg-$Mg_2Ca$ interface in the centre of the pillar. As shown in Figure 6b, the slip event at the interface can be characterized as a non-dissociated prismatic <a> dislocation (See also Movie S6 for the MD counterpart).

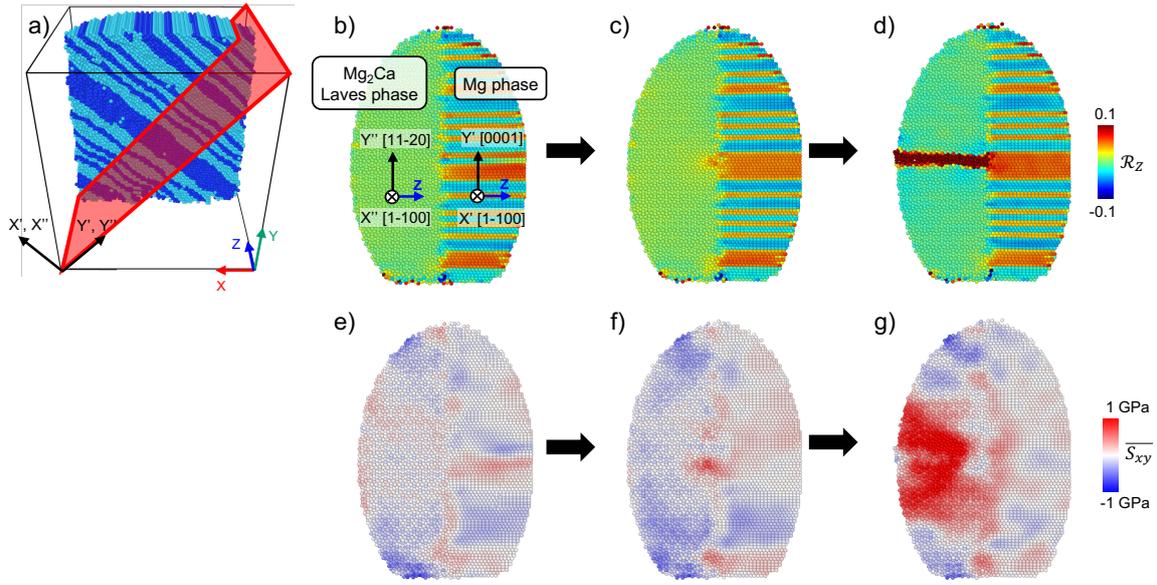

Figure 7: Kinematics of the onset of plasticity at the Mg-Mg$_2$Ca interface during nanomechanical compression tests with the orientation relationship $(0001)_\alpha \parallel (11\bar{2}0)_{Mg_2Ca}$. **a)** perspective view similar to Figure 6a, showing the plane visualized in b) to g). The atoms in the Laves phase are removed for clarity. **b-g)** sliced profile views of the onset of Laves phase plasticity, corresponding to the half-transparent red plane of the perspective view on the left. All atoms visible. **b,c,d)** atoms are coloured according to $\mathcal{R}_Z$, the local rigid-body rotation around the Z axis (arbitrary unit) highlighting dislocations and twins. **e,f,g)** atoms are coloured according to $\overline{S_{xy}}$, the averaged (over a sphere of 10 Å) local atomic shear stress (in GPa).

The driving force for the onset of plasticity in the Laves phase is unveiled in Figure 7 showing sliced views normal to the glide planes at the onset of Laves phase plasticity. Figure 7b-d represents the evolution of slip by the calculation of the local rigid-body rotation $\mathcal{R}_Z$ around the Z axis. Local rotation is especially useful to highlight twining. Figure 7e-g represents the evolution of to $\overline{S_{xy}}$, the averaged (over the neighbours within a sphere of 10 Å) local atomic shear stress. In particular, the nucleation of the dislocation in the Laves phase is revealed by an orange spot at the interface (Figure 7c) and the fully propagated dislocation is clearly evidenced by the dark red band (Figure 7d). Note that the nucleation of the dislocation in the Laves phase correlates to the propagation of one <a> partial dislocation in the Mg phase (Figure 7b,c). The nucleation of the dislocation in the Laves phase is directly related to a stress concentration at the Mg-Mg$_2$Ca interface (Figure 7f) that released the stress initially stored in the Mg phase (Figure 7e). While releasing the global stress of the pillar (at 15% strain, as shown in *Figure 5*b), the propagation of the dislocation in the Laves phase generates additional shear stresses in this phase of the order of 1 GPa (Figure 7g), favouring the subsequent plasticity of the intermetallic phase.

### 3.2.3. Simple shear of dislocation pile-up

To gain detailed and well-defined insights into the mechanisms of interfacial plasticity, it is common to use carefully controlled two dimensional atomistic setups [36–38]. Such approaches include infinite straight dislocations with controlled character, semi-infinite planar interfaces and controlled stress states. In our work, we consider a quasi-2D bi-crystal slab made of the Mg phase and the C14-Mg$_2$Ca phase with the OR $(0001)_\alpha \parallel (0001)_{Mg_2Ca}$ and the OR $(0001)_\alpha \parallel (11\bar{2}0)_{Mg_2Ca}$. Figure 8a shows the Mg-Mg$_2$Ca atomistic composite sample, with periodic boundary condition (BC) along the X and Y axes. The OR $(0001)_\alpha \parallel (11\bar{2}0)_{Mg_2Ca}$ is chosen as an example, but we observed identical results for the OR $(0001)_\alpha \parallel (0001)_{Mg_2Ca}$. 2D BC are applied along the Z axis (red layers in Figure 8a). From 1 to 20 infinite straight basal edge dislocations are inserted in the centre of the simulation within the Mg phase. The dislocations are successively inserted by using Atomsk [30]. The system is relaxed by MS (see section 3.1) and deformed by quasi-static simulations as described in the following. A strain $\varepsilon_{yz}$ is applied on the sample by homogeneously displacing all the atoms, followed by a complete relaxation of the system by MS. To maintain the deformation during the minimization, a force $F$ is applied in the Y direction on each atom that belong to the 2D BC, and defined by $F = (\varepsilon_{yz}.\overline{\mu^*})/(S.N)$ with $\overline{\mu^*}$ being the average shear modulus of the sample for the particular orientations, $S$ the surface area along the Z direction and $N$ the number of atoms in the 2D BC.

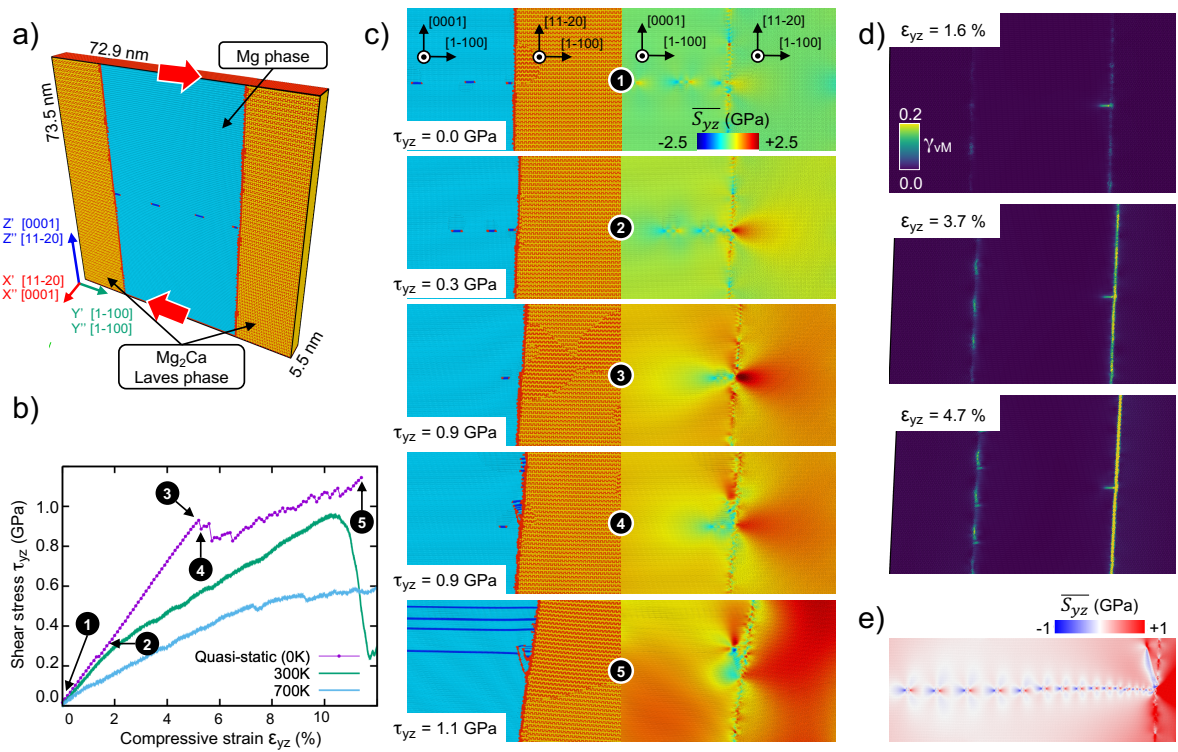

*Figure 8: Dislocation pile-up at the Mg-Mg$_2$Ca interface with orientation relationship $(0001)_\alpha \parallel (11\bar{2}0)_{Mg_2Ca}$. **a)** setup of the quasi-static simulations. Red arrows indicate the shear direction. Atoms are coloured according to the common neighbour analysis (CNA): perfect hexagonal crystal, light blue; stacking fault in hex, dark blue; C14 Laves phase, orange and red; other structure type, red only. **b)** stress-strain curves of the quasi-static (0K) and finite temperature (300K and 700K) shear deformations. The numbers in the black circles refer to the snapshots in c). Each dot on the curve indicates a relaxed state. **c)** snapshots of the quasi-static deformation for different global shear stresses, τ, as referred in b). Left colouring: CNA. Right colouring: local shear stress $\overline{S_{yz}}$. **d)** interfacial sliding revealed by snapshots at different strain level during the linear regime. Colour according to the atomic shear strain γ$_{vM}$. **e)** pile-up of 20 dislocations at an applied shear stress of approximately 1 GPa, highlighted by the local shear stress $\overline{S_{yz}}$.*

A non-sheared relaxed sample containing four dislocations is shown in Figure 8a. Two of the four dislocations are in contact with the interface. The stress-strain profile of the quasi-static deformation shown in Figure 8b (purple line) exhibits a linear stress increase up to $\tau = 0.9$ GPa. This corresponds to the absorption of the first three Mg dislocations, as evidenced in Figure 8c revealing a sequence of dislocation pile-ups interacting with the interface (left views highlight the crystallographic structure by CNA, right views evidence the local atomic shear stress). At an applied shear stress $\tau = 0.9$ GPa, the interface nucleates defects (Figure 8c) that release not only the global stress (Figure 8b) but also the stress induced into the $Mg_2Ca$ phase by the dislocation absorbed at the interface (see Figure 8c, snapshot 3 and 4). Due to this stress release, no defect is transmitted or formed in the $Mg_2Ca$ phase. However, the nucleated interfacial defects continue to grow, while additional defects form and propagate into the Mg phase (Figure 8c (5), dark blue bands). We conclude that the interface fails after the absorption of four dislocations. It is interesting to notice that during the linear stage of stress increase, the interface with absorbed dislocations undergoes significant local sliding (Figure 8d, right vertical interface with a strong increase of local atomic strain $\gamma_{vM}$). On the contrary, the interface that did not absorb any dislocations from the pile-up shows negligible local sliding (Figure 8d, left vertical interface with a low increase of local atomic strain $\gamma_{vM}$).

The behaviour we report above is independent of the size of the dislocation pile-up and of the orientation relationship. The maximum number of dislocations the interface can absorb is three. As an example, Figure 8e shows a pile-up of 20 dislocations in front of a failed interface, 3 dislocations being already absorbed in the interface.

We investigated the influence of temperature on the behaviour we report above. The two dimensional set up with a pile-up of four dislocations has been deformed at finite temperatures within the canonical ensemble (NVT) at a shear rate of $10^8$ s$^{-1}$. Figure 8b shows the stress-strain curves of these deformation tests at 300K and 700K in green and light blue, respectively. Similar to the quasi-static simulations, the stress drops correspond to the nucleation of defects in the Mg phase or to the failure of the interface. No transfer of slip from the Mg to the $Mg_2Ca$ phase is observed (See also Movie S7). The behaviour presented above appears thus independent of the temperature. A comprehensive investigation of the effect of the temperature on the brittle to ductile transition in such composite is however out of the scope of this work.

### 3.3. Discussion on the simulations

The numerical deformation tests we present in this work show slip transmission across the Mg-$Mg_2Ca$ interface. We identified two regimes depending on the orientation relationship. When the basal planes of the matrix are aligned with the basal planes of the Laves phase, a direct slip transmission can occur. It consists of the accumulation of matrix <a> basal full dislocations at the interface, that directly trigger the nucleation of a partial dislocation on the basal planes of the Laves phase. The temperature dependency indicates that this mechanism is thermally activated, similar to the propagation of the synchroshear dislocation in bulk Laves phases [6,35,39]. Note that a synchroshear dislocation propagates by the synchronous glide of two ordinary Shockleys on adjacent parallel atomic planes, as described in detail elsewhere [6,39,40]. On the other hand, when the basal planes of the matrix are aligned with the prismatic planes of the Laves phase, an indirect transmission can occur. It consists of the local accumulation of shear stress within the interface plane, the resulting torsion indirectly triggering the nucleation of a full non-dissociated <a> dislocation on the prismatic plane of the Laves phase. This mechanism seems thus to be strongly facilitated by deformation twinning in the matrix. Note that the in-depth analysis of dislocation mechanisms in such complex atomistic structures requires novel technics that remain to be developed.

These two regimes for slip transmission can be, however, entirely inhibited by interfacial sliding. In particular, when the applied load favoured sliding within planes parallel to the interface plane, local sliding of the atoms at the interface appears to release sufficient internal stress to prevent the nucleation of dislocations in the Laves phase.

## 4. Discussion on experimental and simulation results

In this work, experimental and numerical approaches were combined to investigate the plasticity and co-deformation of Mg-Laves phase composites. As discussed in the previous sections, the stress required to activate plasticity in the Mg phase is significantly lower than that required in case of the Laves phases. Consequently, under the application of external loading, the Mg phase deforms first and then at the stress concentration points, three different events can take place: i) cracking in the Laves phase, ii) slip transfer from α-Mg to Laves phase and iii) interfacial sliding at the α-Mg-Laves phase interfaces. The details pertaining to each event are covered separately in the following sections.

### 4.1. Cracking in the Laves phase

The atomistic simulation results (*Figure 3*) confirm that under application of tensile stress plasticity is first initiated in the α-Mg phase and then further straining results in the initiation of fracture in the Mg$_2$Ca phase especially at low temperatures, i.e. at 50 K. This happens preferably at places where basal slip lines in α-Mg matrix interacts with the interfaces. Due to limited ductility, experimental observations of pronounced deformation are largely limited to higher temperatures, but the same combination of cracking and slip events in the adjacent Laves phase have indeed been found also in experiments after deformation at ≈170°C (see *Figure 1* and [22]).

### 4.2. Slip transfer from α-Mg to Laves phase

Alternatively, simulations also confirm that, at the same stress concentration points (where basal slip lines in the α-Mg matrix interacts with the Laves phase), plasticity may be transmitted to the Laves phase, especially at elevated temperatures (500K and 700 K, see *Figure 3* and *Figure 4*). We further showed that the basal <a> slip in the α-Mg matrix activates plasticity in the Laves phase within the basal planes. Similar results were again also obtained in our experiments, where plasticity from the α-Mg phase can be seen transmitted to the basal planes of the Laves phase (**Error! Reference source not found.**) at the stress concentration points.

In some cases, as for our simulation setup with the orientation relationship $(0001)_\alpha \parallel (11\bar{2}0)_{Mg_2Ca}$, the slip event in the Mg₂Ca activated from the interface was based on non-dissociated prismatic <a> dislocations. In agreement with this, observations of prismatic slip during nano-indentation have also been made by Zehnder et al. [2]. In fact, the CRSS for 1$^{st}$ order prismatic slip (≈440 MPa) was reported to be lower compared with basal slip (≈520 MPa) at room temperature [2].

### 4.3. Interfacial sliding at the α-Mg-Laves phase interfaces

Our atomistic results suggest that interface sliding can prevent the transfer of matrix slip to the Laves phase. Interfacial sliding is primarily achieved by the absorption of dislocations at the interface. This is in agreement with earlier experimental observations, where interfacial sliding of Mg-Laves phase interfaces has been reported [22,41].

However, unfortunately we have not been able to provide an experimental correlation between slip transfer or local interfacial sliding and the local stress state and orientation relationship, as the number of events we observed remained limited. Nonetheless, our simulations show that the slip transfer mechanism depends on the interfacial orientation and can be inhibited for particular geometries, even at high temperature (see Figure 8). This highlights the importance of considering local strain distributions and local interfacial geometries to predict slip transfer at phase boundaries. Experimentally, these insights may help guide future experiments to target specific orientation relationships and to investigate the thermal activation of interface-dominated deformation by nanomechanical testing. In this way, simulations and experiments will directly benefit from each other to unravel the co-deformation mechanisms of the two phases.

## 5. Conclusions

In this work, we combined experimental and numerical approaches to investigate the plasticity and co-deformation of Mg-Laves phase composites. From this work we conclude that:

- The mechanisms of co-deformation, namely cracking, slip transmission and interfacial sliding as a result of slip activity in the Mg phase, are observed in both simulation and experiment.

- For slip transfer, the Laves phase deforms on those planes identified previously in experiments and simulations, the prismatic and basal planes. The simulations suggest that this should occur by synchroshear on the basal plane and non-dissociated <a> dislocation on the prismatic plane.
- In the case of interfacial sliding, we found that dislocations are absorbed into the interface in the simulations. An insight that may be validated experimentally by targeted deformation and transmission electron microscopy in the future.
- Overall, our simulations reveal that the active co-deformation mechanism depends on the interfacial orientation. This highlights the importance of considering local strain distributions and local interfacial geometries to predict slip transfer at phase boundaries.

More work has to be performed to understand the full extent of slip transfer at such complex interfaces. In particular, our numerical results suggest that this transfer might be thermally activated, but this remains to be confirmed by dedicated experiments. The findings we present here on the transfer of slip across Laves phase interfaces in Mg alloys pave the way for in-depth investigations of the plasticity in such complex lightweight composite. More generally, understanding how the plasticity is transferred at phase boundaries opens up exploration paths for composites with tailored mechanical properties.

## Acknowledgments

The authors acknowledge financial support by the Deutsche Forschungsgemeinschaft (DFG) through the projects A02, A03, A05, C01 and C02 of the SFB1394 Structural and Chemical Atomic Complexity – From Defect Phase Diagrams to Material Properties, project ID 409476157. Simulations were performed with computing resources granted by RWTH Aachen University (rwth0297, rwth0407, rwth0591 and prep0017) and by the GENCI-TGCC (grant 2020-A0080911390).

## Data availability

The raw/processed data required to reproduce these findings cannot be shared at this time as the data also forms part of an ongoing study.

Foundry. 14 (2017) 265–271. https://doi.org/10.1007/s41230-017-7077-z.